\documentclass[prd,superscriptaddress,amsfonts,amssymb,amsmath,showpacs,twocolumn,showkeys]{revtex4-1}

\usepackage{amsmath}
\usepackage{epsfig}
\usepackage{graphics}
\usepackage{color}
\usepackage{graphicx}
\usepackage[colorlinks=true,
      linkcolor=black,
      urlcolor=red,
      citecolor=blue]{hyperref}

\begin{document}
\title{Light deflection by charged wormholes in Einstein-Maxwell-dilaton theory}
\author{Kimet Jusufi}
\email{kimet.jusufi@unite.edu.mk}
\affiliation{Physics Department, State University of Tetovo, Ilinden Street nn, 1200,
Tetovo, Macedonia}
\affiliation{Institute of Physics, Faculty of Natural Sciences and Mathematics, Ss. Cyril and Methodius University, Arhimedova 3, 1000 Skopje, Macedonia}

\author{Ali \"{O}vg\"{u}n}
\email{ali.ovgun@pucv.cl}

\affiliation{Instituto de F\'{\i}sica, Pontificia Universidad Cat\'olica de
Valpara\'{\i}so, Casilla 4950, Valpara\'{\i}so, Chile}

\affiliation{Physics Department, Arts and Sciences Faculty, Eastern Mediterranean University, Famagusta, North Cyprus via Mersin 10, Turkey}

\author{Ayan Banerjee}
\email{ayan\_7575@yahoo.co.in}
\affiliation{Department of
Mathematics, Jadavpur University, Kolkata 700 032, West Bengal,
India}
\date{\today }

\begin{abstract}
In this paper, we study the deflection of light by a class of charged wormholes within the context of the Einstein-Maxwell-dilaton theory. The primordial wormholes are predicted to exist in the early universe, where inflation driven by the dilaton field. We perform our analysis through optical geometry using the Gibbons-Werner method (GW), by adopting the Gauss-Bonnet theorem and the standard geodesics approach. We report an interesting result for the deflection angle in leading-order terms--namely, the deflection angle increases due to the electric charge $Q$ and the magnetic charge $P$, whereas it decreases due to the dilaton charge $\Sigma$. Finally, we confirm our findings by means of geodesics equations. Our computations show that the GW method gives an exact result in leading order terms.

\end{abstract}

\maketitle

\section{Introduction}
An important finding of Einstein's theory of relativity is that light rays are deflected by a gravitational field. Einstein calculated that the deflection predicted by his theory would be twice the Newtonian value. During a total solar eclipse in 1919, Eddington performed the first experimental test of GR \cite{edd}.  Gravitational lensing is a powerful tool of astrophysics and cosmology that can be used to measure the mass of galaxies and clusters, and to detect dark matter \cite{dm}. Now, a century later, we have calculated the deflection angle of light deflected by charged wormholes (CW) within the context of the Einstein-Maxwell-dilaton (EMD) theory, using the Gauss-Bonnet theorem (GBT). 

Since the Big Bang, the universe has been expanding and cooling down while remaining uniform and isotropic.  There is a phase transition in the cooling time, related to the breaking of the symmetry, that causes topological defects. It is assumed that inflation is driven by the scalar field, namely the inflaton \cite{ujp}. Moreover, typical dilaton fields are also quite suitable  for producing the correct value of the slow-roll of the inflation. On the other hand, string cosmologists believe they can solve this issue by using the kinetic part of a dilaton field and it causes the universe to expand from a flat, cold, and weakly coupled unstable initial vacuum state toward a curved, dilaton-driven, strong coupling regime which is called the pre-Big-Bang phase \cite{nojiri,gdil,kuntz,hendi}. Furthermore, the solutions of the classical black holes and wormholes can exist in the development of EMD theory \cite{Goulart1}.

The wormholes solutions represent a shortcut between the points of two parallel universes or, two different points of the same universe. Those objects are among the most intriguing and intensively studied topics in general relativity \cite{MM,MV}. In brief, traversable wormholes are supported 
within the context of general relativity by matter with stress-energy tensor that violates 
the null energy condition and, according to which, 
exotic matter is required in order to keep throat of the wormholes open. Actually, the wormhole solutions violate all the energy conditions \cite{MV}. The idea of a wormhole can be traced back to Flamm, who first proposed the wormhole idea in 1916, just after the discovery of Schwarzschild's black hole solution. Then, in 1935 Einstein and Rosen introduced a bridge-like structure between black holes (today known as Einstein-Rosen bridge) in order to obtain a regular solution without any singularities \cite{tw2}. However, the term wormhole was coined by the Wheeler in 1957 \cite{WH,FW}. The modern interest in a traversable wormhole was stimulated particularly by the pioneering works of Morris, Thorne and Yurtsever \cite{MM,MTY}. Traversable wormholes in this sense are described as having throats that 
connect two asymptotically flat regions of spacetime. Discussions revolve around the physical conditions required for traversable wormholes within the context of general relativity.
Nowadays, the most challenging problem in classical gravitational physics is to construct a traversable and a stable wormhole solution with ordinary matter. There are well-known examples of wormholes such as classical, minimally coupled, massless scalar field and electric charges, as reported in the literature \cite{lobo,Sushkov,garattini}. Moreover, Goulart \cite{Goulart2} has recently obtained zero-mass point-like solutions that arise from the dyonic black hole solution of EMD theory. This shows 
that from the nonextremal solution, it is possible to construct a static CW solution that satisfies the null energy condition.

Nowadays, the applications of the strong/weak gravitational lensing by wormholes are a very active area of research. From an observational point of view, gravitational lensing is an important window into the probing of wormholes, and the trajectory of an Ellis wormhole light ray was investigated by Ellis in \cite{Ellis}. Afterwards, the deflection angle of the Ellis wormhole spacetime has been calculated by Chetouani and Clement \cite{wh0}. In a recent paper, Tsukamoto and Harada \cite{wh11} have studied the light curve of a light ray that passes through the throat of a traversable wormhole. Additionally, they showed that gravitational lensing can be used as a probe to distinguish between wormholes and black holes \cite{wh12,wh13}. Due to importance of this problem, various articles, such as \cite{wh1,wh2,wh3,wh4,wh6,CB,wh8,wh9,wh10}, have studied gravitational lensing by wormholes.

Gibbons and Werner recently showed, using the example of light deflection from a Schwarzschild de-Sitter black hole \cite{gibbons1} that the angle of light deflection can be calculated using the GBT. This method relies on the fact that the deflection angle can be calculated using a domain outside of the light ray. It is known that the effect of lensing strongly depends on the  mass of the enclosed region body on spacetime. The GBT simply glues surfaces together. One must first use the Euler characteristic of $\chi$ and a Riemannian metric of $g$. The subset-oriented surface domain of $(D,\chi,g)$ is chosen to calculate the Gaussian curvature of $K$, so that the GBT is found as follows:
\begin{equation}
\int\int_D K \rm d S + \int_{\partial D} \kappa \rm d t +\sum_i \alpha_i = 2\pi \chi(D).
\label{gb}
\end{equation} Here, $\kappa$ stands for the geodesics curvature of $\partial D:\{t\}\rightarrow D$, and $\alpha_i$ is the exterior angle with the $i^{th}$ vertex. 

Werner extended the GBT to the stationary spacetimes by given the example of the Kerr deflection angle to the osculating Riemannian metric \cite{werner}. This technique is for asymptotically flat observers and sources. The resulting deflection angle is expected to be too small, which is also a joint point in astronomy. In this paper, we will use the extended geometric method; similar to the works of Gibbons and Werner. In this method, Riemannian metric manifolds are global symmetric lenses. First, we will calculate the Gaussian curvature of $K$ in an optical geometry to obtain the asymptotic deflecting angle of alpha:
\begin{equation}
\hat{\alpha}=-\int \int_{D_\infty} K \mathrm{d}S.
\end{equation}

Note that we use the infinite region of the surface $D_\infty$ bounded by the light ray to calculate our integral. To obtain the deflection angle of the light, we use the zero-order approximation of the light ray, and the deflection angle of $\hat{\alpha}$ is obtained in leading-order terms. Following seminal papers of Gibbons and Werner, many other studies appeared in the literature such as the deflection angle in spacetimes with topological defects, including cosmic strings and global monopoles, quantum effects on the deflection of light by quantum-improved Schwarzschild black holes \cite{kimet1,kimet2}, and gravitational effects due to a cosmic string in Schwarzschild spacetime \cite{kimet3}. Moreover, using the GBT on the Rindler-modified Schwarzschild black hole, Sakalli and \"{O}vg\"{u}n recently showed the deflection angle at the infrared region \cite{aovgun}. The method of calculating the bending angle of light using the GBT has been extensively studied in \cite{asahi1,asahi2,asahi3}.

In this present work, our goal is to apply the
GBT to calculate the angle of light deflection by massless
CW. The paper is organized as follows: in Section II, we review the wormhole solution as presented in EMD theory. In Section III, we consider the deflection angle of light in a CW geometry in the weak-limit approximation using the GBT. In Section IV, we explore geodesics equations to recover the deflection angle. Finally, we draw our conclusions in Section V. Throughout this paper, we will be using natural units, i.e., $G=c=\hbar=1$.

\section{Dyonic Wormholes in the Einstein-Maxwell-dilaton theory}

In this section, we review briefly the solution of the dyonic wormholes in the EMD theory \cite{Goulart1,Goulart2}. Let us firstly consider the simplest action, which can be written as follows
(adopting geometrised units and henceforth, $16\pi G =1$):

\begin{equation}
S=\int d^{4}x\sqrt{-g}\left(R-2\partial_{\mu}\phi\partial^{\mu}\phi-W(\phi)F_{\mu\nu}F^{\mu\nu}\right),\label{ad}
\end{equation}
where R denotes the Ricci scalar, $\phi$ is the dilaton scalar field, and $F_{\mu \nu}$ represents the electromagnetic
field strength, which is given by
\begin{equation}
F_{\mu\nu}=\partial_{\mu}A_{\nu}-\partial_{\nu}A_{\mu}.
\end{equation}
 
Now, considering the equations of motion for the metric, dilaton, and gauge fields, and then Bianchi identities arising from the action (3):

\begin{equation}
R_{\mu\nu}=2\partial_{\mu}\phi\partial_{\nu}\phi-\frac{1}{2}g_{\mu\nu} W(\phi)F_{\rho\sigma}F^{\rho\sigma}+2W(\phi)F_{\mu\rho}F_{\nu}^{\rho},
\end{equation}

\begin{equation}
\nabla_{\mu}\left(\partial^{\mu} \phi\right)-\frac{1}{4}\frac{\partial W(\phi)}{\partial \phi}F_{\mu\nu}F^{\mu\nu} =0,
\end{equation}
\begin{equation}
\nabla_{\mu}\left(W(\phi)F^{\mu\nu}\right) =0,
\end{equation}
\begin{equation}
\nabla_{[\mu}{F_{\rho\sigma]}} =0.
\end{equation}

For the sake of generality, we shall consider $W(\phi)=e^{-2\phi}$ \cite{Goulart1}, following this, 
we obtain the bosonic sector $SU(4)$ version of $N = 4$, per the supergravity theory for a constant axion field.
A doubly charged black hole solution has been found in the bosonic sector of $N = 4$, $d = 4$ 
supergravity \cite {MR}, given a static, axially symmetric spacetime. Moreover, rotating dyonic black holes of $N = 4$, $SO(4)$-gauged supergravity have been considered in \cite{AD}.
Now, we are interested in a dyonic black hole of $N = 4$ to the $SU(4)$ supergravity theory in terms of the integration constants, so we choose a spherically symmetric metric, expressed as \cite{Goulart1}: 
 
\begin{equation}
\mathrm{d}s^{2}=-f(r)\mathrm{d}t^{2}+\frac{1}{f(r)}\mathrm{d}r^{2}+h(r)\mathrm{d}\Omega_2^2, \label{genmet}
\end{equation}
where $\mathrm{d}\Omega_2^2 = \mathrm{d}\theta^{2}+\sin^{2}\theta \mathrm{d}\phi^{2}$ denotes the line element of the unit 2-sphere and the metric functions are 
\begin{align}
f(r) & =\frac{(r-r_{1})(r-r_{2})}{(r+d_{0})(r+d_{1})},\,\,\,h(r)=(r+d_{0})(r+d_{1}),\label{genmets}\\
e^{2\phi} & =e^{2\phi_{0}}\frac{r+d_{1}}{r+d_{0}},\label{gendil}\\
F_{rt} & =\frac{e^{2\phi_{0}}Q}{(r+d_{0})^{2}},\,\,\,F_{\theta\phi}=P\sin\theta,\label{genele}
\end{align}
where $P$ is the magnetic charge, $Q$ is the electric charge, the value of the dilaton at infinity is
$\phi_0$, with four integrating constant $r_1$, $r_2$, $d_0$ and $d_1$.

Now, we are interested in a massless pointlike dyonic solution. Nevertheless, 
for this purposes we consider the case when $d_{1}= -d_{0}= -\Sigma$ and 
$r_{1}= -r_{2} \equiv r_{H}$, to get the
following relation \cite{Goulart1}
\begin{equation}
e^{2\phi_{0}} =\pm \frac{P}{Q}.
\end{equation}

As suggested by Goulart in \cite{Goulart1}, we only consider the case of a negative sign, 
as a positive solution does not correspond to the massless solution obtained by EMD theory. In this physical situation, considering the negative sign in Eq. (13), the nonextremal
solutions from Eq.s (10-12) can be explicitly written as:
\begin{align}
f(r) & =\frac{(r-r_{+})(r-r_{-})}{(r^{2}-\Sigma^{2})},\,\,h(r)=(r^{2}-\Sigma^{2}), \label{metKal}\\
e^{2\phi} & =-\frac{P}{Q}\left(\frac{r-\Sigma}{r+\Sigma}\right),\\
F_{rt} & = -\frac{P}{(r+\Sigma)^{2}},\,\,\,F_{\theta\phi}=P\sin\theta.
\end{align}

The horizon and singularity are located at
\begin{equation}
r_{+}= +\sqrt{\Sigma^2+2QP}, ~~~r_{S} = |\Sigma|, 
\end{equation}
which excludes the inner horizon, and the area of the two-sphere shrinks to zero
at $r_{S}$. The main importance of this solution is that the massless solution seems physically acceptable,
even with a complex dilaton field at infinity. 

Given the above results and the full massless nonextremal
solution (14), one must choose the negative sign in Eq. (13) with the constants
$d_{1}= -d_{0}= -\Sigma$. Following this method, the obtained metric is \cite{Goulart2}:
\begin{equation} 
\begin{split}
\mathrm{d}s^2& = -\left(\frac{r^2}{r^2+2PQ}\right)\mathrm{d}t^2+\left(\frac{r^2+2PQ}{r^2+\Sigma^2+2PQ}\right)\mathrm{d}r^2 \\
 & +(r^2+2QP)(\mathrm{d}\theta^2+\sin^2\theta \mathrm{d}\varphi^2).
\end{split}
\end{equation}

The above metric represents the CW in the EMD theory, 
which can be obtained from massless nonextremal dyonic solutions. It is worth noting that by letting $\Sigma=0$, the radius of the throat is found to be $\mathcal{R}_{thro.}=\sqrt{2PQ}$ \cite{Goulart2}.

Next, we study the deflection of light produced by  
a CW geometry within the context of the EMD theory.

\section{Weak deflection limit with GBT }
\subsection{Gaussian Optical Curvature}

Let's use the Goulart's wormhole solution given in Eq. (18) by considering the null geodesic  $\mathrm{d}s^{2}=0$, with the deflection angle of light in the equatorial plane $\theta =\pi /2$, we obtain the optical metric of CW as follows:
\begin{equation}
\mathrm{d}t^{2}=\frac{(r^2+2PQ)^2}{r^2(r^2+\Sigma^2+2PQ)}\mathrm{d}r^2+\frac{(r^2+2PQ)^2}{r^2}\mathrm{d}\varphi^2.
\end{equation}

Now, we introduce a radial Regge-Wheeler tortoise
type coordinate $r^{\star }$, with a new function $%
f(r^{\star })$ as follows
\begin{equation}
\mathrm{d}r^{\star }=\frac{r^2+2PQ}{r\sqrt{r^2+\Sigma^2+2PQ}}\mathrm{d}r,\,\,\,f(r^{\star })=\frac{(r^2+2PQ)}{r}.
\end{equation}

Then, the above metric becomes
\begin{equation}
\mathrm{d}t^{2}=\tilde{g}_{ab}\,\mathrm{d}x^{a}\mathrm{d}x^{b}=\mathrm{d}{%
r^{\star }}^{2}+f^{2}(r^{\star })\mathrm{d}\varphi ^{2}. \,\,\,\,(a,b=r,\varphi).
\end{equation}

The above optical metric has two nonzero Christoffel symbols
\begin{eqnarray}
\tilde{\Gamma} _{\varphi \varphi }^{r}&=&\frac { 4P^2Q^2-r^4}{r^3}
,\\
\tilde{\Gamma} _{r\varphi }^{\varphi}&=& \frac{r^2-2PQ}{r\left(r^2+2PQ\right)}.
\end{eqnarray} 

Note that we have used the approximation $\mathrm{d}r^{\star}\approx \mathrm{d}r$ in the last two equations, which is convenient in our setup for a very large $r$. It is straightforward to compute the Gaussian optical curvature $K$, which can be calculated by the following equation \cite{gibbons1}: 
\begin{eqnarray}
K &=&-\frac{1}{f(r^{\star })}\frac{\mathrm{d}^{2}f(r^{\star })}{\mathrm{d}{%
r^{\star }}^{2}} \\
&=&-\frac{1}{f(r^{\star })}\left[ \frac{\mathrm{d}r}{\mathrm{d}r^{\star }}%
\frac{\mathrm{d}}{\mathrm{d}r}\left( \frac{\mathrm{d}r}{\mathrm{d}r^{\star }}%
\right) \frac{\mathrm{d}f}{\mathrm{d}r}+\left( \frac{\mathrm{d}r}{\mathrm{d}%
r^{\star }}\right) ^{2}\frac{\mathrm{d}^{2}f}{\mathrm{d}r^{2}}\right]. \notag
\end{eqnarray}

Using the Eq. (21) the Gaussian optical curvature for Goulart's wormhole gives
\begin{widetext}
\begin{equation}
K=-\frac{6 PQ r^4 +8 PQ \Sigma^2 r^2-r^4 \Sigma^2+8 P^3 Q^3+16 P^2 Q^2 r^2+4P^2 Q^2 }{\left(2PQ+r^2\right)^4}.
\end{equation}
\end{widetext}
Since we are interested in the weak limit, we can approximate the optical Gaussian curvature as 
\begin{equation}\label{Curvature}
K\approx -\frac{16 PQ}{r^4}+\frac{\Sigma^2 }{r^4}-\frac{16 PQ \Sigma^2}{r^6}+\frac{32 P^2 Q^2}{r^6}.
\end{equation}

Later on, we shall use this important result together with the GBT to find
the deflection angle in the following section.

\subsection{Deflection angle}

Having calculated the Gaussian optical curvature, we use this relationship and apply the GBT to the optical geometry of the Goulart's wormhole. Let us choose a non-singular region $\mathcal{D}_{R}$ with boundary $\partial
\mathcal{D}_{R}=\gamma _{\tilde{g}}\cup C_{R}$, which allows the GBT to be stated as follows \cite{gibbons1}: 
\begin{equation}
\iint\limits_{\mathcal{D}_{R}}K\,\mathrm{d}S+\oint\limits_{\partial \mathcal{%
D}_{R}}\kappa \,\mathrm{d}t+\sum_{i}\theta _{i}=2\pi \chi (\mathcal{D}_{R}),
\end{equation}
in which $\kappa $ gives the geodesic curvature, $K$ stands for the Gaussian optical curvature, while $\theta_{i}$ is the exterior angle at the $i^{th}$ vertex. It is seen from Fig.1, that we can choose a non-singular domain outside of the light ray with the Euler characteristic number $%
\chi (\mathcal{D}_{R})=1$.

\begin{figure}[h!]
\includegraphics[width=0.47\textwidth]{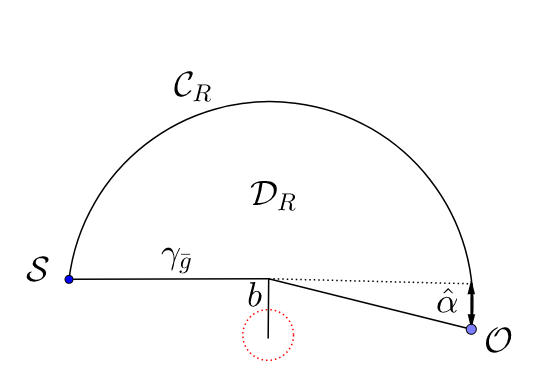} 
\caption{\small \textit {Deflection angle of light in the wormhole geometry in the equatorial plane $(r,\varphi)$. In our set-up, $b$ is the impact parameter and can be approximated as the distance of the closest approach $r_{min}$ of the light trajectory from the coordinate origin located at the center of the wormhole. The radius of the throat $\mathcal{R}_{thro.}$ is negligible compared to the impact parameter $b$.}}
\end{figure}

In order to find the deflection angle, let us first compute the geodesic curvature using the following relation
\begin{equation}
\kappa =\tilde{g}\,\left(\nabla _{\dot{%
\gamma}}\dot{\gamma},\ddot{\gamma}\right) 
\end{equation}
together with the unit speed condition $\tilde{g}(\dot{\gamma},\dot{%
\gamma})=1$, where $\ddot{\gamma}$ gives the unit acceleration vector. If we let $R\rightarrow \infty $, our two jump angles ($\theta _{\mathcal{O}}$, $%
\theta _{\mathcal{S}}$) become $\pi /2,$ or in other words, the sum of jump angles to the source $\mathcal{S}$, and observer $\mathcal{O}$, satisfies $\theta _{\mathit{O}%
}+\theta _{\mathit{S}}\rightarrow \pi $ \cite{gibbons1}. Hence we can write GBT as
\begin{equation}
\iint\limits_{\mathcal{D}_{R}}K\,\mathrm{d}S+\oint\limits_{C_{R}}\kappa \,%
\mathrm{d}t\overset{{R\rightarrow \infty }}{=}\iint\limits_{\mathcal{D}%
_{\infty }}K\,\mathrm{d}S+\int\limits_{0}^{\pi +\hat{\alpha}}\mathrm{d}\varphi
=\pi.
\end{equation}

Let us now compute the geodesic curvature $\kappa$. To do so, we first point out that $\kappa (\gamma _{\tilde{g}})=0$, since $\gamma _{\tilde{g}}$ is a
geodesic. We are left with the following
\begin{equation}
\kappa (C_{R})=|\nabla _{\dot{C}_{R}}\dot{C}_{R}|,
\end{equation}
where we choose $C_{R}:=r(\varphi)=R=\text{const}$. The radial part is evaluated as
\begin{equation}
\left( \nabla _{\dot{C}_{R}}\dot{C}_{R}\right) ^{r}=\dot{C}_{R}^{\varphi
}\,\left( \partial _{\varphi }\dot{C}_{R}^{r}\right) +\tilde{\Gamma} _{\varphi
\varphi }^{r}\left( \dot{C}_{R}^{\varphi }\right) ^{2}. \label{12}
\end{equation}

From the last equation, it  obvious that the first term vanishes, while the second term is calculated using Eq. (23) and the unit speed condition. For the geodesic curvature we find
\begin{eqnarray}\notag
\lim_{R\rightarrow \infty }\kappa (C_{R}) &=&\lim_{R\rightarrow \infty
}\left\vert \nabla _{\dot{C}_{R}}\dot{C}_{R}\right\vert , \notag \\
&=&\lim_{R\rightarrow \infty }\left( \frac{R^2-2PQ}{R^3+2PQR}\right), \notag \\
&\rightarrow &\frac{1}{R}. 
\end{eqnarray}%

On the other hand, for very large radial distance Eq. (20) yields
\begin{eqnarray}\notag
\lim_{R\rightarrow \infty } \mathrm{d}t &=&\lim_{R\rightarrow \infty
}\left(\frac{R^2+2PQ}{R}\right) \mathrm{d}\varphi\\
&\to & R \, \mathrm{d}\varphi.  
\end{eqnarray}%

If we combine the last two equations, we find $
\kappa (C_{R})\mathrm{d}t= \mathrm{d}\,\varphi
$. It is convenient to choose the deflection line as $r=b/\sin \varphi$, in that case, the deflection angle from Eq. (30) can be recast in the following from 
\begin{eqnarray}\label{int0}
\hat{\alpha}=-\int\limits_{0}^{\pi}\int\limits_{\frac{b}{\sin \varphi}}^{\infty}K\mathrm{d}S.
\end{eqnarray}

If we substitute Eq. \eqref{Curvature} into the last equation,  this yields the following integral
\begin{widetext}
\begin{eqnarray}\label{int1}
\hat{\alpha}&=&-\int\limits_{0}^{\pi}\int\limits_{\frac{b}{\sin \varphi}}^{\infty}\left(-\frac{16 PQ}{r^4}+\frac{\Sigma^2 }{r^4}-\frac{16 PQ \Sigma^2}{r^6}+\frac{32 P^2 Q^2}{r^6}\right)\sqrt{\det \tilde{g}}\,\mathrm{d}r\mathrm{d}\varphi.
\end{eqnarray}
\end{widetext}

Note that we use the following relation $\mathrm{d}r^{\star} \approx \mathrm{d}r$, valid in the limit as $R\to \infty $. One can easily solve this integral in the leading order terms to find the following result
\begin{equation}\label{GB1}
\hat{\alpha}\simeq \frac{3 \pi P Q}{2 b^2}-\frac{\pi \Sigma^2}{4 b^2}+\mathcal{O}(P^2,Q^2,\Sigma^2).
\end{equation}

It is worth noting that we use a straight-line approximation to evaluate the integral \eqref{int1}; therefore, we expect that only the first-order terms should be valid in our setup. However, the Eq. \eqref{int0} gives an exact expression for the deflection angle when integrated over the domain $\mathcal{D}_\infty$. But, in principle, if we use an appropriate equation for the light ray $r$ which includes higher order terms of $P$, $Q$, and $\Sigma$, one should recover the second-order correction terms by carrying out the integration over the domain $\mathcal{D}_\infty$.

\section{Geodesic Equations}

In this section, we further show that one can indeed reach the same result \eqref{GB1} by using the standard geodesic approach. To do so, we recall that the variational principle stated as follows:
\begin{equation}
\delta \int \mathcal{L} \,\mathrm{d}s=0.
\end{equation}

When we apply it to our wormhole spacetime metric (18), the Lagrangian reads:
\begin{widetext}
\begin{eqnarray}\label{geo1}
2\,\mathcal{L}=-\frac{r^2(s)\dot{t}^2(s)}{r^2(s)+2PQ}+\frac{\left(r^2(s)+2PQ\right)\dot{r}^2(s)}{r^2(s)+\Sigma^2+2PQ}+\left(r^2(s)+2PQ\right)\dot{\varphi}^2(s).
\end{eqnarray}
\end{widetext}

Note that we have three cases, namely $\mathcal{L}$ being $+1,0,-1$, for timelike, null, and spacelike geodesics, respectively.
Without loss of generality, we consider the deflection of planar photons i.e. $\theta =\pi/2$. After using the spacetime symmetries, one should consider two constants of motion $l$ and $\gamma$, given as follows \cite{Boyer}:
\begin{eqnarray}
p_{\varphi}&=&\frac{\partial \mathcal{L}}{\partial \dot{\varphi}}=2(r^2(s)+2PQ)\dot{\varphi}(s)=l,\\
p_{t}&=&\frac{\partial \mathcal{L}}{\partial \dot{t}}=-\frac{2 \,r^2(s)\dot{t}(s)}{r^2(s)+2PQ}=-\gamma.
\end{eqnarray}

Let us now introduce a new variable $u(\varphi)$, which is related to our old radial coordinate as follows $r=1/u(\varphi)$ and hence we obtain the following identity:
\begin{equation}\label{iden}
\frac{\dot{r}}{\dot{\varphi}}=\frac{\mathrm{d}r}{\mathrm{d}\varphi}=-\frac{1}{u^2}\frac{\mathrm{d}u}{\mathrm{d}\varphi}.
\end{equation}

Without loss of generality we can normalize the affine parameter along the light rays by taking $\gamma=1$ \cite{Boyer} and approximate the distance of closest approach with the impact parameter i.e., $u_{max}=1/r_{min}=1/b$, since we shall consider only leading order terms \cite{lorio}. In this case, one can choose the second constant of motion as follows:
\begin{equation}
l=\sqrt{\frac{\Xi(P,Q)}{2PQ+b^2}}\,b^2,
\end{equation}
where
\begin{equation}
\Xi(P,Q)=\frac{12 P^2 Q^2}{b^4}+\frac{6 PQ}{b^2}+1.
\end{equation}

We see from the last two equations that if we take the limit $P=Q=\Sigma=0$, then $l=b$. Finally using Eqs. (39), (42), (43), in terms of $u(\varphi)$, we find the following equation:
\begin{widetext}
\begin{eqnarray}\label{33}
\left(\frac{\mathrm{d}u}{\mathrm{d}\varphi}\right)^2\left[\frac{2PQu^2+1}{\left(2PQ+\Sigma^2+\frac{1}{u^2}\right)u^6} \right]+\frac{1}{u^2}-\frac{u^2 \left(\frac{2PQ}{b^2}+1\right)\left( \frac{1}{u^4}+\frac{4PQ}{u^2}+4P^2 Q^2 \right)^2 }{\Xi\, b^2 \left(2PQ+\frac{1}{u^2} \right)}+2PQ=0.
\end{eqnarray}
\end{widetext}

One way to solve this equation is to use a perturbation method. Note that by setting $P=Q=\Sigma^2=0$, and then performing a differentiation yields the expected results
\begin{equation}
\frac{\mathrm{d}^2u_0}{\mathrm{d}\varphi^2}+u_0=0.
\end{equation}

It is well known that the solution of the differential equation (45) is given by the following relation \cite{Boyer,weinberg}
\begin{equation}
\Delta \varphi =\pi+\hat{\alpha},
\end{equation}
where $\hat{\alpha}$ is the deflection angle to be calculated. Following the same arguments given in Ref. \cite{weinberg}, the deflection angle can be calculated as
\begin{equation}
\hat{\alpha}=2|\varphi(u_{max})-\varphi_{\infty}|-\pi.
\end{equation}
where 
\begin{equation}
\varphi=\int_0^{1/b} \mathcal{A}(P,Q,\Sigma^2,u) \,\mathrm{d}u.
\end{equation}

Note that in the last equation $\mathcal{A}(P,Q,\Sigma^2,u)$ is calculated by considering Taylor expansion series around $Q$, $P$, and $\Sigma^2$, given by
\begin{widetext}
\begin{eqnarray}\notag
\mathcal{A}(u)&=& -{\frac {15\,{b}^{2} \left( bu-1 \right) ^{2} \left( bu+1 \right) ^{2}
}{4\,\sqrt {-{b}^{8}{u}^{2}+{b}^{6}} \left( {b}^{2}{u}^{2}-1 \right) ^
{2}}} \Big[ \left( -{\frac {4}{15}}+{P}^{2}{Q}^{2}{u}^{6}{\Sigma}^{2}-\frac{2PQ
}{5} \left( PQ+{\Sigma}^{2} \right) {u}^{4}+ \left( {\frac {4\,PQ}{15}}+\frac{2\Sigma^2}{15} \right) {u}^{2} \right) {b}^{2}\\
&-& \frac{1}{15}\, \left( 8\,PQ{u}^{4}{\Sigma}^{2}-4\,{\Sigma}^{2}{u}^{2}+8
 \right) QP \Big].
\end{eqnarray}
\end{widetext}

As expected, the deflection angle in the weak limit approximation is found to be the same result found by GBT
\begin{equation}
\hat{\alpha}\simeq \frac{3 \pi P Q}{2 b^2}-\frac{\pi \Sigma^2}{4 b^2}+\mathcal{O}(P^2,Q^2,\Sigma^2).
\end{equation}

As we have pointed out, these equations agree only for the first-order terms while the agreement between these methods breaks down for the second-order correction terms. 

\section{Conclusion}

In this paper, we studied gravitational lensing by a CW geometry within the context of the EMD theory. Adopting the weak deflection limit, we calculated the deflection angle, and found the deflection angle is affected by the magnetic charge, electric charge, and the dilaton charge. In particular, the magnetic and electric charges increase the deflection angle. On the other hand, the dilaton charge decreases the deflection angle. To obtain these results, we used two different approaches: the GW method and geodesic equations. In the first method, we have applied the GBT to optical geometry of the Goulart's wormhole in the equatorial plane. We first calculated the Gaussian optical curvature by integrating over a nonsingular domain outside the light ray. The first important finding is that the GW method gives an exact result in leading-order terms; whereas the second important result emphasizes the role of global topology in the lensing effect. 

In addition, it is now known from current observations that the universe began at extremely high temperatures. This is called the hot Big Bang model. When the universe expanded adiabatically at an accelerated rate, it cooled down, and as a consequence of cosmological phase transitions in the early universe  cosmological defects were produced \cite{kible1,kible2,weak1}. These cosmological defects may lead to the formation of wormholes. The nature of the wormhole formations that occurred in the early universe may be detected using weak lensing observations. The relationship between weak lensing and quasinormal modes (QNMs) is also important because it extends the theorem of Hod \cite{hod} to the theories of Einstein-Gauss-Bonnet and shows that in the WKB limit, there is an universal upper bound
for the real part of the QNMs. On the other hand, in the strong lensing regime, there is an universal lower bound on black holes \cite{weak4}. Studying of weak gravitational lensing also provide possible evidences for the validity of the Cosmic Censorship conjecture (CCC) \cite{werner2}. Moreover, the astrophysical importance of gravitational lensing and geodesics studies has given us an interest in working on a proof of the Hod's conjecture for wormholes \cite{hod}. We will leave for future work the consideration of the second-order correction terms in the GW method in the weak lensing limit, as well as the strong limit with QNM in the wormhole spacetimes.

\begin{acknowledgments}
This work was supported by the Chilean FONDECYT Grant No. 3170035 (A\"{O}). AB is thankful to the authority of Inter-University Centre for Astronomy and Astrophysics, Pune, India for providing research facilities. 
\end{acknowledgments}

\end{document}